\documentclass[aps,prl,twocolumn,showpacs,preprintnumbers,superscriptaddress]{revtex4}

\usepackage{graphicx}
\usepackage{epsf}
\usepackage{dcolumn}
\usepackage{bm}
\usepackage{amssymb}
\usepackage{amsmath}

\begin{document}

\newcommand{\be}{\begin{eqnarray}}
\newcommand{\ee}{\end{eqnarray}}

\title{Anomalous Roughening of Viscous Fluid Fronts in Spontaneous Imbibition}

\author{J. Soriano}
\affiliation{Experimentalphysik I, Universit\"{a}t Bayreuth. Universit\"{a}tstra\ss e 30.
D-95447 Bayreuth, Germany}

\author{A. Mercier}
\affiliation{\'Ecole Normale Sup\'erieure. All\'ee d'Italie 46.
F-69364 Lyon, France}

\author{R. Planet}
\affiliation{Departament d'Estructura i Constituents de la Mat\`eria, Universitat de
Barcelona, Av.\ Diagonal 647, E-08028 Barcelona, Spain}

\author{A. Hern\'andez--Machado}
\affiliation{Departament d'Estructura i Constituents de la Mat\`eria,
Universitat de Barcelona, Av.\ Diagonal 647, E-08028 Barcelona,
Spain}

\author{M.A. Rodr{\'\i}guez}
\affiliation{Instituto de F\'{\i}sica de Cantabria (IFCA), CSIC--UC,
E-39005 Santander, Spain}

\author{J. Ort\'{\i}n}
\email{ortin@ecm.ub.es}
\affiliation{Departament d'Estructura i Constituents de la
Mat\`eria, Universitat de Barcelona, Av.\ Diagonal 647, E-08028 Barcelona, Spain}

\date{\today}

\begin{abstract}

We report experiments on spontaneous imbibition of a viscous fluid by a model porous
medium in the absence of gravity. The average position of the interface satisfies
Washburn's law. Scaling of the interface fluctuations suggests a dynamic exponent $z
\simeq 3$, indicative of global dynamics driven by capillary forces. The complete set of
exponents clearly shows that interfaces are not self-affine, exhibiting distinct local
and global scaling, both for time ($\beta = 0.64 \pm 0.02$, $\beta^* = 0.33 \pm 0.03$)
and space ($\alpha = 1.94 \pm 0.20$, $\alpha_{loc} = 0.94 \pm 0.10$). These values are
compatible with an intrinsic anomalous scaling scenario.

\end{abstract}


\maketitle

The dynamics of immiscible fluid--fluid displacements in porous media has been a subject
of much interest in last years
\cite{Alava,Sahimi-95,Barabasi-Stanley,Zik-97,Rubio-89,Soriano-2002-I,Soriano-2002,Soriano-2002-II,Geromichalos-2002},
both from a fundamental point of view, as a dynamical non equilibrium process leading to
rough interfaces \cite{Alava}, and from a technological point of view, in industrial and
environmental problems such as oil recovery, irrigation, and filtration \cite{Sahimi-95}.
The process is called {\it spontaneous imbibition} when an invading fluid that wets
preferentially the medium displaces a resident fluid at constant external pressure. In
contrast, {\it forced--flow imbibition} takes place when the invading fluid gets into the
porous medium at constant injection rate.

The case of spontaneous imbibition has special interest from the point of view of dynamic scaling. Spontaneous
imbibition is dominated by capillarity. In contrast to forced--flow imbibition, there is no global mass
conservation law, a fact that may have a dramatic effect on scaling \cite{Dube1999,Dube2001}. Spontaneous
imbibition is a case with slowing--down dynamics in which time scales change continuously. Indeed, it is known
from a long time \cite{wash} that the temporal scaling of the mean advancing front obeys Washburn's law, $\langle
h \rangle \sim t^{1/2}$, but little is known about the scaling of fluctuations in this regime. Moreover, most
experiments in spontaneous imbibition have been performed with the disordered medium placed vertically
\cite{Geromichalos-2002}, a situation where gravity limits the Washburn behavior to a short observation time.
Experiments with gravity have been directed to the study of the pinning process and its scaling. Only one
experiment with the disordered medium placed horizontally has been performed on paper \cite{Zik-97}. In this
experiment scale invariant roughness is found only for highly anisotropic disorder, while logarithmic roughness,
at best, is found in isotropic paper. Paper, however, is a medium with uncontrolled disorder, and phenomena such
as swelling and prewetting might change the effect of static capillary forces.

In this Letter we report on spontaneous imbibition experiments of a viscous fluid by a
model porous medium with well controlled disorder, consisting on a horizontal Hele--Shaw
cell with random variations in gap spacing. Our scaling results suggest a dynamic
exponent $z \simeq 3$, consistent with a nonlocal dynamics driven by capillary forces.
This exponent has never been observed in experiments before, despite having been
suggested by simple models of imbibition \cite{Alava,Aurora01}. After a detailed scaling
analysis of the interfacial fluctuations we find different global and local exponents,
with values that suggest that our experiment must be described in the framework of
intrinsic anomalous scaling \cite{generic}.

\paragraph{Experimental setup.--}
The experimental setup \cite{Soriano-2002-I,Soriano-2002} consists on a Hele--Shaw cell
of size $190 \times 500$ mm$^2$ (L $\times$ H) made of two glass plates separated a
narrow gap of thickness $b=0.46$ mm. Fluctuations in the gap spacing are provided by
copper obstacles ($d=0.06$ mm high, $1.50 \times 1.50$ mm$^2$ size) that are randomly
distributed over a fiberglass substrate, without overlap, filling 35\% of the substrate
area (disorder SQ 1.50 in Ref.\ \cite{Soriano-2002-I}).

\begin{figure}
\centerline{\epsfxsize=8cm \epsfbox{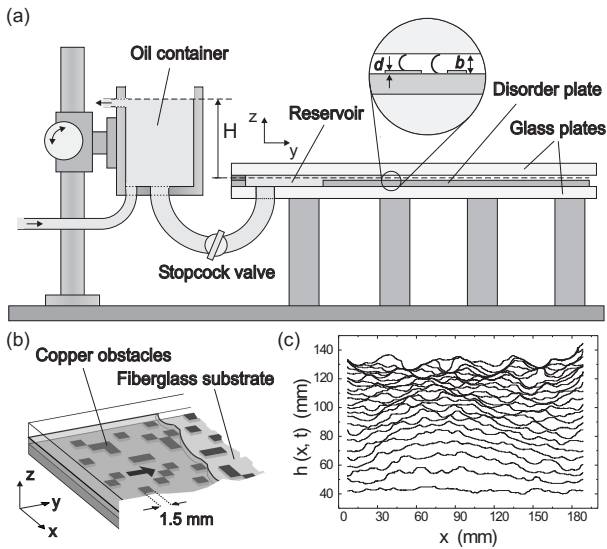}} \caption{(a) Sketch of the experimental
setup. (b) Schematic representation of our Hele-Shaw cell with quenched disorder in the
bottom plate. (c) Temporal evolution of the oil-air interface at $H = -9$ mm.}
\label{Fig:Setup}
\end{figure}

In the present set of experiments we use a constant pressure device that consists on an
oil container of selectable height $H$, in the range from $-50$ to $100$ mm, in steps of
$0.1$ mm. A silicone oil (Rhodorsil 47 V) with kinematic viscosity $\nu = 50$ mm$^2$/s,
density $\rho = 998$ kg/m$^3$, and oil--air surface tension $\sigma = 20.7$ mN/m at room
temperature, penetrates into the cell through two wide holes drilled in the bottom glass,
displacing the air initially present. The holes connect to a reservoir in the cell, with
the disorder plate placed just ahead (Fig.\ \ref{Fig:Setup}).

The experimental procedure used to prepare a flat initial interface is the following: The
reservoir in the cell is carefully filled and, using an auxiliary syringe, the oil--air
interface is pushed gently up to a transverse copper track 2 mm behind the disorder
pattern, and the experiment starts. The evolution of the oil--air interface is monitored
using two CCD cameras. A computer records the images acquired and stores them for
processing.

\begin{figure}
\centerline{\epsfxsize=7.5cm \epsfbox{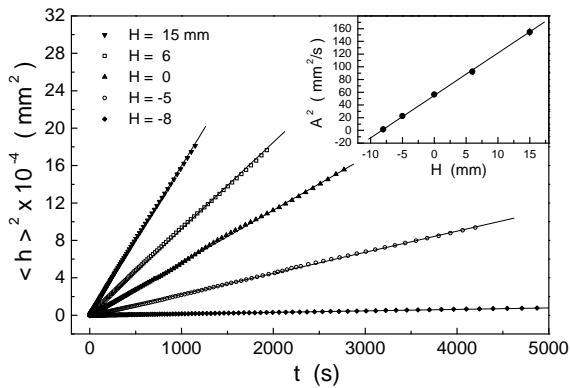}} \caption{Main plot:
Squared average interface position as a function of time, for
different oil column heights $H$, showing that the imbibition process
follows Washburn's law, $\langle h \rangle = A t^{1/2}$. Inset:
Dependence of Washburn's law prefactor $A$ (squared) on column height
$H$.} \label{Fig:Washburn}
\end{figure}

\paragraph{Washburn's law.--}
Different heights of the oil column in the container, in the range $-9 \leq H \leq 15$
mm, have been explored. One example of the temporal evolution of the interface is shown
in Fig.\ \ref{Fig:Setup}(c) for $H=-9$ mm. As a first result (Fig.\ \ref{Fig:Washburn})
we observe that for $H > -10$ mm the average interface position obeys Washburn's law,
$\langle h \rangle = A t^{1/2}$. For $H < -10$ mm the interfaces recede. By Darcy's law
$\langle \dot{h} \rangle \sim \overrightarrow{\nabla} p \sim H/ \langle h \rangle$, so
that $A^{2} \sim H$, as verified in the inset of Fig.\ \ref{Fig:Washburn}.

\paragraph{Scaling analysis of the rough fronts.--}
The statistical properties of a one--dimensional interface defined by a function $h(x,t)$
are usually described in terms of the fluctuations of $h$. More precisely, {\em global}
fluctuations are measured by the global interface width, which for a system of total
lateral size $L$ scales according to the Family-Vicsek (FV) {\em ansatz} \cite{Family-Vicsek-1985}.
However, in experiments one cannot usually probe the global width, since system size is
fixed in standard setups. Instead of global fluctuations, one measures the {\em local}
surface fluctuations, which are more easily accessible in the laboratory. Local interface
fluctuations are measured by calculating either the spectral density of the interface,
$S(k,t) \equiv \langle h(k,t)h(-k,t)\rangle$, where $h(k,t)$ is the Fourier transform of
the interface and brackets indicate averages over independent runs of the experiment, or
the local width, $w(l,t) = \langle \langle [h(x,t) - \langle h
\rangle_l]^2\rangle_l\rangle^{1/2}$, where $\langle \cdots \rangle_{l}$ denotes an
average over $x$ in windows of size $l$. The local width is also expected to satisfy
dynamic scaling
\begin{equation}
\label{anom-scal} w(l,t) = t^{\beta} g(l/t^{1/z}),
\end{equation}
where the most general scaling function \cite{lopez97,generic} is
expected to behave as
\begin{equation}
\label{g} g(u) \sim \left\{ \begin{array}{lcl}
    u^{\alpha_{loc}} & {\rm if} & u \ll 1 \\
    \mbox{const.} & {\rm if} & u \gg 1
\end{array}
\right..
\end{equation}
Here $\beta$ and $z$ are the growth and dynamic exponents,
respectively, and $\alpha = \beta z$ is the roughness exponent.
So, for short scales ($l<t^{1/z}$) the local width grows as
$t^{\beta^*}$ with $\beta^* = (\alpha-\alpha_{loc})/z$, whereas
for larger scales the growth of the global width, $t^{\beta}$, is
recovered. This generalization of the scaling form is required to
account for potentially {\em anomalous} roughening ($\alpha \neq
\alpha_{loc}$). Note that three independent exponents are
necessary now to define the universality class. The standard
self--affine FV scaling \cite{Family-Vicsek-1985} is then fully
recovered when
{\em global} and {\em local} roughness exponents become equal,
i.e. $\alpha = \alpha_{loc}, \beta^* =0$. Sometimes a more
detailed analysis of the scaling functions is necessary to
completely define the generic class of scaling. Then both
$S(k,t)$ and $w(l,t)$ must be examined. The scaling of the spectrum
\begin{equation}
\label{spectrum} S(k,t)= k^{-(2\alpha+1)} s(kt^{1/z}),
\end{equation}
where now the scaling function is
\begin{equation}
\label{s} s(u) \sim \left\{ \begin{array}{lcl}
    u^{2(\alpha-\alpha_{s})} & {\rm if} & u \gg 1 \\
    u^{2\alpha+1} & {\rm if} & u \ll 1
\end{array}
\right.,
\end{equation}
and $\alpha_{s}$ is the {\it spectral} exponent, adds valuable additional information
\cite{generic}. With the same value of the scaling exponents the system can belong to one
or another class depending on the value of the auxiliary exponent $\alpha_{s}$. For
instance, with $\alpha>1$ the system can be either in the superrrough,
$\alpha_{s}=\alpha$, or the intrinsic anomalous class, $\alpha_{s}=\alpha_{loc}$. This is
important because it can inform of the existence of different symmetries \cite{lopez05}.

\paragraph{Experimental results.--}

Since slight perturbations of the interface at the beginning of the experiment are
unavoidable, it is difficult to characterize the scaling of the interfacial fluctuations
at very short times. To minimize this effect we have always considered the subtracted
width $W(l,t)$, defined as $W(l,t)=[w^2(l,t) - w^2(l,0)]^{1/2}$ \cite{Barabasi-Stanley}.

\begin{figure}
\centerline{\epsfxsize=7.5cm \epsfbox{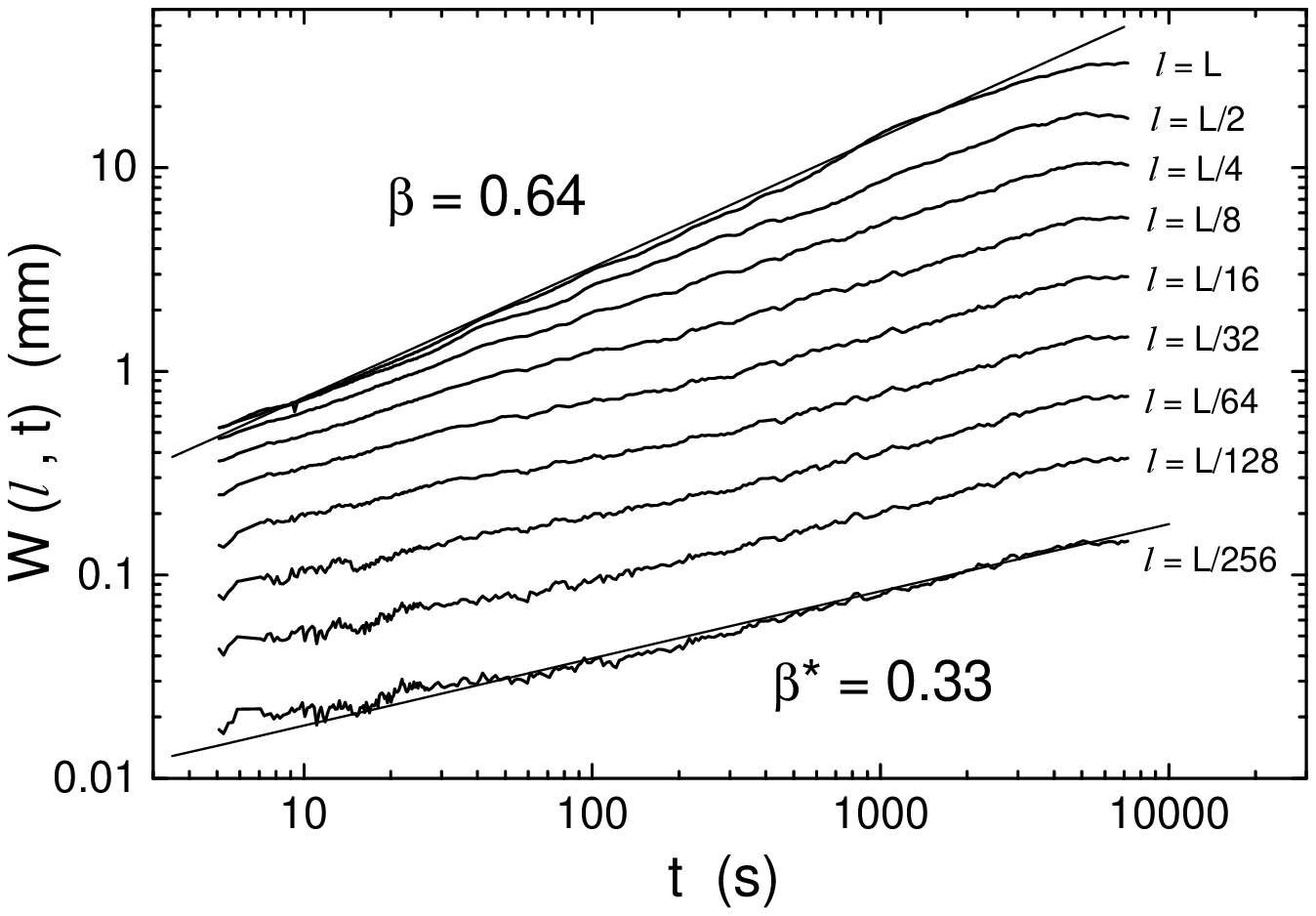}} \caption{Log--log
plot of the interfacial width as a function of time, for different window sizes $l$.
The experimental parameters are $b=0.46$ mm and $H=-9$ mm.
The straight lines, with slopes $0.64$ and $0.33$, result from a data fit in the scaling region.}
\label{Fig:width}
\end{figure}

We have focused on measuring the values of the scaling exponents for an oil column height
$H=-9$ mm. In these conditions the imposed (negative) pressure gradient nearly
compensates the traction due to capillarity, so that the interface is driven very slowly.
The prefactor of Washburn's law in this case is $A = 1.26 \pm 0.15$ mm/s$^{1/2}$, so that
the average interface velocity (initially $\langle \dot{h} (t=0) \rangle
\simeq 0.05$ mm/s) decreases by about 84\% in
the experiment. We have performed a total of $6$ different experiments, $2$ for each of
$3$ different disorder realizations. The plot of $W(l,t)$ as a function of time and for 9
different windows of size $l$ is shown in Fig.\ \ref{Fig:width}. We obtain a power law
with different exponents for large and small window size, $\beta = 0.64 \pm 0.02$ and
$\beta^* = 0.33 \pm 0.03$. The value of this last exponent is very robust and clearly
different from $0$. Saturation occurs simultaneously in all scales at around $6800$ s. Both
facts rule out a FV scaling and indicate the presence of anomalous scaling.

\begin{figure}
\centerline{\epsfxsize=7.5cm \epsfbox{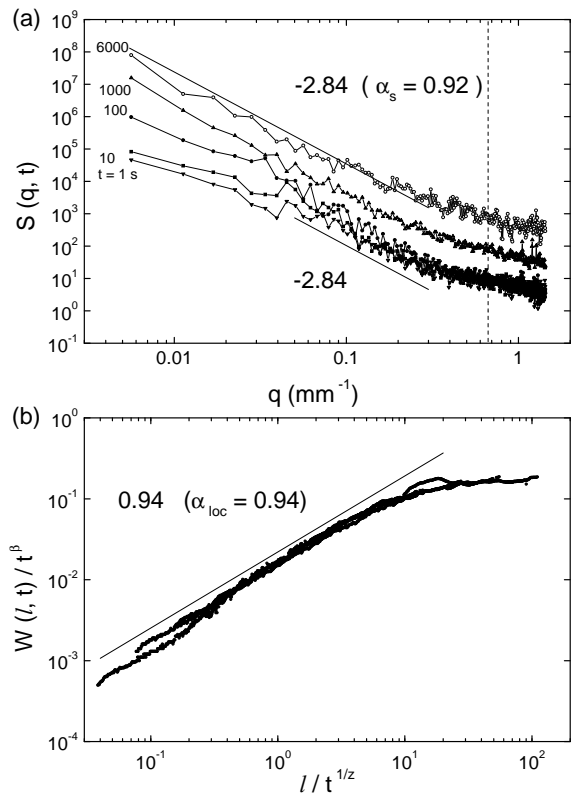}} \caption{(a) Temporal evolution of the
power spectra of the interfaces analyzed in Fig.\ \ref{Fig:width}. The solid straight
lines, with slope $-2.84$, are fits to the data in the scaling regions. The vertical
dashed line indicates the value $q = 0.67$ associated with the lateral size of the unit copper
obstacle. (b) Collapse of the interfacial width shown in Fig.\ \ref{Fig:width}, using the
values $\beta = 0.64$ and $z = 3$. The straight line is the result from a data fit.}
\label{Fig:power}
\end{figure}

The temporal evolution of the power spectrum, presented in Fig.\ \ref{Fig:power}(a), provides the value $\alpha_s
= 0.92 \pm 0.11$. In addition, the observation that the spectra at different times are shifted upwards rules out
the possibility of superroughness. The remaining scenario, $\alpha_{loc}=\alpha_s$, can be checked up on the
collapse suggested in Eq.\ (\ref{anom-scal}). The result of the best collapse, obtained with $\beta = 0.64$
(previously determined) and $z=3$, is shown in Fig.\ \ref{Fig:power}(b). From this collapse we obtain
$\alpha_{loc} = 0.94 \pm 0.10$ as a result of a data fit. Having determined the three independent exponents
$\beta$, $\beta^*$ and $\alpha_{loc}$ directly from the experimental data, the remaining exponents $\alpha = 1.94
\pm 0.20$ and $z = 3.0 \pm 0.3$ follow from the scaling relations. We have verified that the collapse of the
spectrum (not shown in the figure) is also well fitted with the exponents thus obtained.

\begin{figure}
\centerline{\epsfxsize=6.5cm \epsfbox{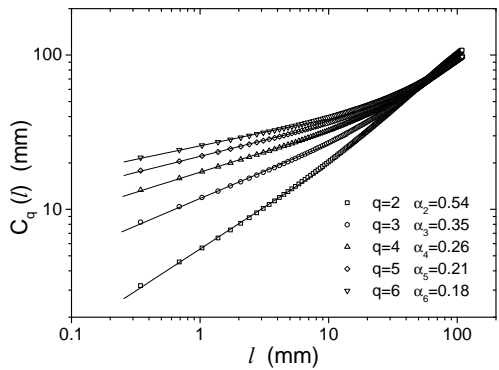}} \caption{Multiscaling analysis performed
through the computation of the q--order height-height correlation function, for interfaces
at saturation.} \label{Fig:multi}
\end{figure}

To provide more evidences of anomalous scaling, we have investigated the possibility of multiscaling, $C_q
(l,t)\equiv \langle [ h(x+l,t)-h(x,t)]^q \rangle ^{1/q} \sim l^{\alpha_q}$. Figure  \ref{Fig:multi} reveals the
existence of a lengthscale separating uniform scaling from multiscaling. Exponents of the multiscaling region
are generally not universal, and depend most of times on particular choices in the experimental arrangement. In
our case multiscaling is probably due to the inhomogeneous local growth in the sharp edge of copper obstacles. The
exponent of the uniform scaling region is universal, and expected to be compatible with $\alpha_{loc}$.
This is reasonably satisfied by our data, as shown in Fig.\ \ref{Fig:multi}.

\paragraph{Discussion.--}

The deterministic linearized equation for the
dynamics of a perturbation of the interface, in the form of a normal
mode of wavenumber $k$, in spontaneous imbibition, reads \cite{Dube1999,Aurora01}:
$\dot{h}_k = - \sigma (b^2 / 12 \mu) k^2 |k| h_k - \langle \dot{h}
\rangle |k| h_k$, where $\mu = \rho \nu$ is the dynamic viscosity
of the fluid, and $\langle \dot{h} \rangle$ is the average
interface velocity. Surface tension (the term proportional to $k^2
|k|$) is responsible for the damping of short wavelength
fluctuations, while viscous pressure (the term proportional to
$\langle \dot{h} \rangle |k|$) damps long wavelength fluctuations.

The two terms, however, cross over at a characteristic length $\xi_{\times} = 2 \pi [\sigma (b^2 / 12\mu)
(1/\langle \dot{h} \rangle) ]^{1/2}$. In spontaneous imbibition, $\langle \dot{h} \rangle \sim t^{-1/2}$, and thus
the crossover length $\xi_{\times} \sim t^{1/4}$. As the interface invades the porous medium, Dub\'e et al.\
\cite{Dube1999} have pointed out that the growth of the lateral correlation length due to the $k^2|k|$ term, $L_c
\sim t^{1/3}$, is effectively dominated by the slower growth of the crossover length, $\xi_{\times} \sim t^{1/4}$. One
should then observe a dynamic exponent $z=4$ for length scales larger than $\xi_{\times}$, a result verified in
Ref.\ \cite{Geromichalos-2002}.

In our experiment, however, $L_c$ never reaches $\xi_{\times}$, and thus the observation of a dynamic exponent $z
\simeq 3$ is fully justified on theoretical grounds. Indeed, (i) at very short times $L_c$ grows from 0 while,
from $\langle \dot{h} (t=0) \rangle$ and the values of $\sigma$, $b$, and $\mu$, we get $\xi_{\times} \simeq 76$
mm, which is about half the cell width. (ii) At saturation ($L_c = L$) $\xi_{\times}$ can be estimated as follows:
$L_c = (Dt)^{1/3}$, where $D \simeq 10^3$ mm$^3$/s is estimated from the inflection in Fig.\ \ref{Fig:power}(b),
so that saturation occurs indeed at a time $t_s = L^3/D \simeq 6800$ s. The interface velocity at $t_s$ can be
determined now from Washburn's law, and then we get $\xi_{\times} \simeq$ 190 mm at saturation, a value that
coincides with the size of the cell. Thus $\xi_{\times}$ is only reached at saturation. (iii) This result is
consistent with the observation that at the end of the experiment $\langle \dot{h} \rangle \sim 0.16 \,\langle
\dot{h}(t=0) \rangle$, and we get again $\xi_{\times} \simeq$ 190 mm. This reflects also that saturation is
reached shortly before the end of the experiment.


\paragraph{Conclusions.--}

Our experiments provide a controlled realization of spontaneous imbibition without gravity. The interface
roughness is a result of the random spatial distribution of capillary forces. Traction due to capillarity is so
strong that the external pressure must be negative to have a very slow (forward) motion of the roughening front.
Once the control parameter (the pressure imposed) is fixed, the front advance verifies Washburn's law.

Our results $\alpha \simeq 2$, $\alpha_{loc} \simeq 1$, $\beta \simeq 2/3$, $\beta^*
\simeq 1/3$ and $z \simeq 3$, together with $\alpha_{s} \simeq \alpha_{loc}$, identify
the present experiment with the so called intrinsic anomalous scaling \cite{generic}.
This type of scaling is entirely different from the usual FV scaling found in the same
setup and with the same kind of disorder, in forced--flow conditions
\cite{Soriano-2002-I}. This substantial difference should be attributed to the absence of
a global mass conservation law.

Finally, our results provide the first experimental evidence of $z \simeq 3$ in spontaneous imbibition. This has
been made possible by driving the interfaces very slowly, thus avoiding the lateral correlation length $L_c$ to
reach the crossover length $\xi_{\times}$ in the experiment. Preliminary experiments in which the
interfaces are driven at much higher pressure ($H = 29$ mm) and, consequently, $\xi_{\times}$ is much smaller
throughout the whole experiment, cannot be properly collapsed with $z=3$.

\begin{acknowledgments}
Fruitful discussions with J.M. L\'opez and M. Pradas are acknowledged. We are also
grateful to A. Comerma and M. Quevedo for technical support.
JS acknowledges the financial support of the European Training Network PHYNECS through
project HPRN-CT-2002-00312, and RP acknowledges a fellowship
from the Direcci\'on General de Investigaci\'on (MEC, Spain). This work is supported by
the DGI through projects BFM2003-07749-C05-02, -C05-03, and BQU2003-05042-C02-02.
\end{acknowledgments}

\end{document}